# From Vision to Validation: A Theory- and Data-Driven Construction of a GCC-Specific AI Adoption Index

**Mohammad Rashed Albous*** MOHAMMAD.ALBOUS@AASU.EDU.KW
*Abdullah Al Salem University, Kuwait*

**Abdel Latef Anouze** A.ANOUZ@QU.EDU.QA
*Qatar University, Qatar*

## ABSTRACT

Artificial intelligence (AI) is rapidly transforming public-sector processes worldwide, yet standardized measures rarely address the unique drivers, governance models, and cultural nuances of the Gulf Cooperation Council (GCC) countries. This study employs a theory-driven foundation derived from an in-depth analysis of literature review and six National AI Strategies (NASs), coupled with a data-driven approach that utilizes a survey of 203 mid- and senior-level government employees and advanced statistical techniques (K-Means clustering, Principal Component Analysis, and Partial Least Squares Structural Equation Modeling). By combining policy insights with empirical evidence, the research develops and validates a novel AI Adoption Index specifically tailored to the GCC public sector. Findings indicate that robust technical infrastructure and clear policy mandates exert the strongest influence on successful AI implementations, overshadowing organizational readiness in early adoption stages. The combined model explains 70% of the variance in AI outcomes, suggesting that resource-rich environments and top-down policy directives can drive rapid but uneven technology uptake. By consolidating key dimensions (Technical Infrastructure (TI), Organizational Readiness (OR), and Governance Environment (GE)) into a single composite index, this study provides a holistic yet context-sensitive tool for benchmarking AI maturity. The index offers actionable guidance for policymakers seeking to harmonize large-scale deployments with ethical and regulatory standards. Beyond advancing academic discourse, these insights inform more strategic allocation of resources, cross-country cooperation, and capacity-building initiatives, thereby supporting sustained AI-driven transformation in the GCC region and beyond.

Authors' Contact Information: Mohammad Rashed Albous, ORCID: 0009-0008-8646-3810, mohammad.albous@aasu.edu.kw, Abdullah Al Salem University, Alkhaldiya, Kuwait; Abdel Latef Anouze, ORCID: 0000-0002-6989-8897, a.anouz@qu.edu.qa, Doha, Qatar University, Qatar.

*Corresponding author.

## 1. Introduction

Artificial Intelligence (AI) has become a central pillar of public-sector innovation around the world, transforming how governments deliver services, manage resources, and engage with citizens. From automating administrative processes to harnessing data analytics for policy development, AI-driven solutions offer the promise of greater efficiency, transparency, and responsiveness in governance. In the Gulf Cooperation Council (GCC) region, this global trend is particularly pronounced. All six GCC member states, Saudi Arabia, the United Arab Emirates (UAE), Qatar, Bahrain, Kuwait, and Oman, have launched or updated National AI Strategies (NASs) between 2018 and 2024, aligning AI development with ambitious national visions for economic diversification and global competitiveness. Landmark frameworks, such as the UAE's National Innovation Strategy and Saudi Arabia's Vision 2030, underscore the region's emphasis on top-down policy mandates and robust investments to accelerate AI adoption (Albous et al., 2025)

Despite the region's robust commitment to digital transformation, it remains challenging to measure AI's actual impact on governance and service delivery in a localized manner. Traditional global indices focusing on e-Government or digital readiness, such as those featured in the UN E-Government Survey (UN DESA, 2024) and the World Bank's GovTech Maturity Index (2022), provide broad benchmarks but often overlook the unique drivers, barriers, and governance models inherent to GCC countries. These tools excel at mapping baseline conditions like internet penetration or basic service digitalization; however, they do not fully capture the AI-specific layers of algorithmic accountability or the top-down policy mandates common in GCC contexts (cf. OECD, 2024).

The impetus for developing a dedicated AI Adoption Index for the GCC thus arises from multiple regional considerations. First, the high levels of ICT infrastructure noted in the ICT Development Index (IDI, 2024) set a unique foundation from which GCC nations can rapidly scale AI solutions. Second, economic diversification goals, reflected in national visions and in the Global Economic Diversification Index (Prasad et al., 2025), underscore AI's role in driving innovation and competitive advantage. Finally, global competitiveness frameworks like the World Economic Forum's Global Competitiveness Report (2020) highlight the growing importance of advanced technologies to national resilience, reinforcing the need for a region-specific tool to measure AI's real-world impact on public governance.

Against this backdrop, the present study adopts a theory- and data-driven framework to create and validate an AI Adoption Index specifically for the GCC public sector. Drawing first on document analysis of related literature and six NASs, we identify the core theoretical dimensions, Technical Infrastructure (TI), Organizational Readiness (OR), and Governance Environment (GE). We then employ data-driven methods, including a survey of mid- and senior-level government employees plus advanced clustering and structural equation modeling, to empirically refine and validate these dimensions. By integrating policy insights with quantitative measures, the resulting index offers a holistic way to assess AI maturity in GCC public institutions.

Ultimately, the proposed index aims to enrich academic discourse on AI in public administration and provide actionable insights for GCC government leaders. By pinpointing strengths and weaknesses in AI deployment, the index can guide resource allocation, support policy refinement, and stimulate targeted capacity-building programs. The next sections first review existing literature on AI adoption and highlight the GCC-specific drivers that justify developing a specialized index. Following this, the paper introduces a conceptual framework that integrates the key theoretical models (such as the Technology Acceptance Model) and region-specific AI policies to form the basis of our index design. We then outline the methodological steps used for data collection and validation, covering survey design, document analysis, and advanced statistical techniques, before presenting our findings and offering practical recommendations for policymakers. This structure lays the groundwork for continuous monitoring and improvement of AI-driven governance in the GCC region.

## 2. A Comprehensive Literature Synthesis

This section synthesizes existing research on AI adoption, digital government transformation, and index development methodologies, with an emphasis on how these insights inform the design of a GCC-specific AI Adoption Index. Organized around five thematic areas, (1) AI in the public sector: global perspectives, (2) e-government and digital maturity frameworks, (3) AI readiness and adoption models, (4) AI adoption in GCC public institutions and the need for a GCC-specific index, and (5) gaps and rationale for a GCC-specific AI Adoption Index, the review examines the global landscape of AI governance, relevant digital maturity models, and the distinct regional dynamics shaping AI initiatives in GCC countries.

### 2.1 AI in the Public Sector: Global Perspectives

Across the globe, AI has become a powerful catalyst for public-sector innovation. According to The AI Index 2024 report by Stanford University, as of April 2024, more than 94 countries have developed or are actively rolling out NASs, illustrating a widespread acknowledgment of AI's transformative potential (Maslej et al., 2024). Building on previous e-government efforts that initially focused on providing basic online services (Wirtz & Müller, 2019), these strategies now harness AI-driven solutions, from interactive chatbots to sophisticated analytics, to streamline service delivery and inform policy decisions. Ongoing assessments, such as the UN E-Government Survey (UN DESA, 2024), underscore the growing shift toward AI-centric governance.

However, research increasingly highlights the need for strong policy frameworks to fully realize AI's benefits, a perspective reinforced by the OECD's Governing with AI report (2024), which emphasizes the importance of institutional readiness and ethical oversight in successfully adopting AI in the public sector. For the GCC specifically, post-NPM governance arrangements and workforce transitions are central lenses for understanding AI rollouts (Albous et al., 2025).

### 2.2 E-Government and Digital Maturity Frameworks

#### 2.2.1 Traditional E-Government Indices

Over the past two decades, multiple indices, such as the United Nations E-Government Development Index (EGDI) and the World Bank's GovTech Maturity Index (2022), have assessed progress in digital governance by measuring variables like online service availability, infrastructure, and human capital. These indices are crucial in providing a high-level snapshot of how well governments leverage digital tools to engage citizens and deliver essential services. However, as underscored by the OECD (2024), existing e-government indices are limited in addressing the deeper technological and organizational complexities of AI-based initiatives, such as machine learning capacity, algorithmic ethics, and real-time data analytics (see also Albous & Alboloushi, 2025).

#### 2.2.2 Digital Maturity Models

In parallel, digital maturity models, originally designed to map progress from basic ICT adoption to more integrated e-services, provide insights into how organizations evolve in their digital transformations (Layne & Lee, 2001). These models typically gauge an entity's readiness across multiple dimensions, such as infrastructure, process integration, and user-centric service design, thereby offering a roadmap for governments and institutions aiming to advance from nascent to fully digitized operations (Valdes et al., 2011; World Bank, 2022).

However, most traditional frameworks focus on generic "digital readiness", overlooking the nuanced technological, ethical, and regulatory requirements inherent in AI deployment (cf. Oxford Insights, 2024). According to OECD (2024), even high-level digital maturity indicators can fail to capture the complexity of AI's legal and ethical dimensions, especially in regions with top-down governance structures. For instance, dimensions like algorithmic transparency, real-time data governance, and ethical guardrails, critical in AI contexts, are often treated as peripheral or assumed to be part of standard ICT readiness (Mustaf et al., 2020). As a result, organizations looking to integrate AI may find these models insufficiently detailed for capturing emerging issues such as machine learning interpretability, workforce retraining needs, and sector-specific regulatory mandates (Bommasani et al., 2021).

Recent attempts at AI-specific maturity models have started to fill this gap, suggesting additional metrics for evaluating data availability, model reliability, and compliance with national or regional AI strategies (Deloitte, 2024; Gartner, 2025). Yet, these remain largely generic and often do not account for top-down governance structures, resource-intensive deployments, and cultural considerations, all factors that can significantly influence AI projects in certain regions, particularly in the GCC (Prasad et al., 2025). Consequently, adapting or extending existing digital maturity models to capture AI-focused ethical, legal, and strategic dimensions remains an urgent task for both scholars and policymakers aiming to ensure responsible AI adoption in the public sector.

## 2.3 AI Readiness and Adoption Models

### 2.3.1 Conceptual and Theoretical Underpinnings

AI-specific adoption research frequently adapts foundational technology acceptance theories, such as the Technology Acceptance Model (TAM; see Figure 1), to account for variables like perceived usefulness and perceived ease of use, which predict users' behavioral intention to adopt a technology (Davis, 1989). In AI contexts, additional considerations around ethics, fairness, transparency, and regulatory compliance have gained prominence (Janssen et al., 2015). Scholars argue that while traditional acceptance models explain user attitudes, they must be expanded to capture the complex interplay among organizational, technological, and policy drivers unique to AI.

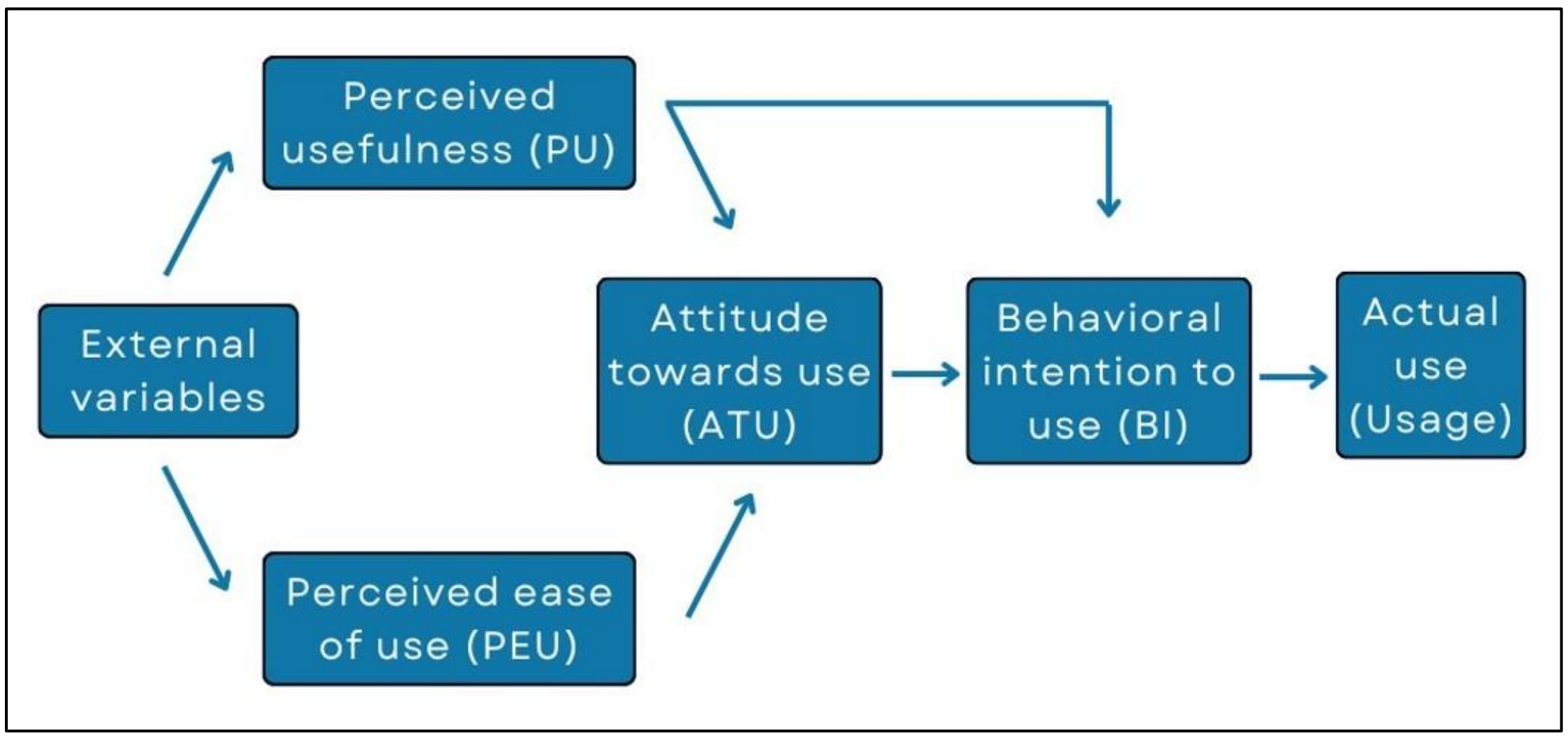

Figure 1. The original technology acceptance model (TAM). Source: Adapted from Davis et al. 1989; Kowalska-Pyzalska 2024.

### 2.3.2 Empirical Studies of AI Adoption

Empirical work underscores the multifaceted nature of AI acceptance across various public domains, including healthcare, education, and social services. Investigations consistently show that success hinges on organizational capabilities, such as staff training and leadership support, and an enabling policy environment (Cameron & Green, 2019; Dwivedi et al., 2021). Recent studies further highlight AI-specific enablers and barriers, including algorithmic transparency, data governance, and workforce retraining programs needed to facilitate widespread adoption (Ki & Kim, 2024; Wirtz et al., 2022).

Importantly, there is also growing recognition that context-specific factors, such as cultural attitudes toward data usage, regulatory clarity, and top-down governance structures, significantly shape AI initiatives in the public sector (Robles & Mallinson, 2025; UNESCO, 2021). For instance, governments with robust legal frameworks and targeted ethical guidelines tend to see higher implementation rates and better public trust in AI-driven services (Laux et al., 2024; OECD, 2024). Conversely, insufficient legislation around data protection or a lack of transparency in algorithmic decision-making can hamper adoption efforts, especially in regions where public scrutiny and stakeholder engagement are evolving. These findings reinforce the notion that effective AI adoption requires not only technical readiness but also supportive institutional norms, ethical oversight, and continuous capacity-building, dimensions often overlooked in generic digital transformation approaches.

## 2.4 AI Adoption in GCC Public Institutions and the Need for a GCC-Specific Index

### 2.4.1 Regional Digital Transformation Agenda

Governments in the GCC, namely Bahrain, Kuwait, Oman, Qatar, Saudi Arabia, and the UAE, are making bold strides in AI development as part of their national strategies for economic diversification and public-sector modernization. Notably, these goals align with indicators tracked by the Global Economic Diversification Index (Prasad et al., 2025), which highlights the growing role of advanced technologies, particularly AI, in driving non-oil growth. Furthermore, between 2018 and 2024, each member state launched or updated a NAS, aligning AI development with broader digital transformation agendas (see Figure 2). These strategies typically emphasize AI for economic diversification and global competitiveness, an approach that resonates with broader national visions such as Saudi Arabia's Vision 2030 and the UAE's National Innovation Strategy.

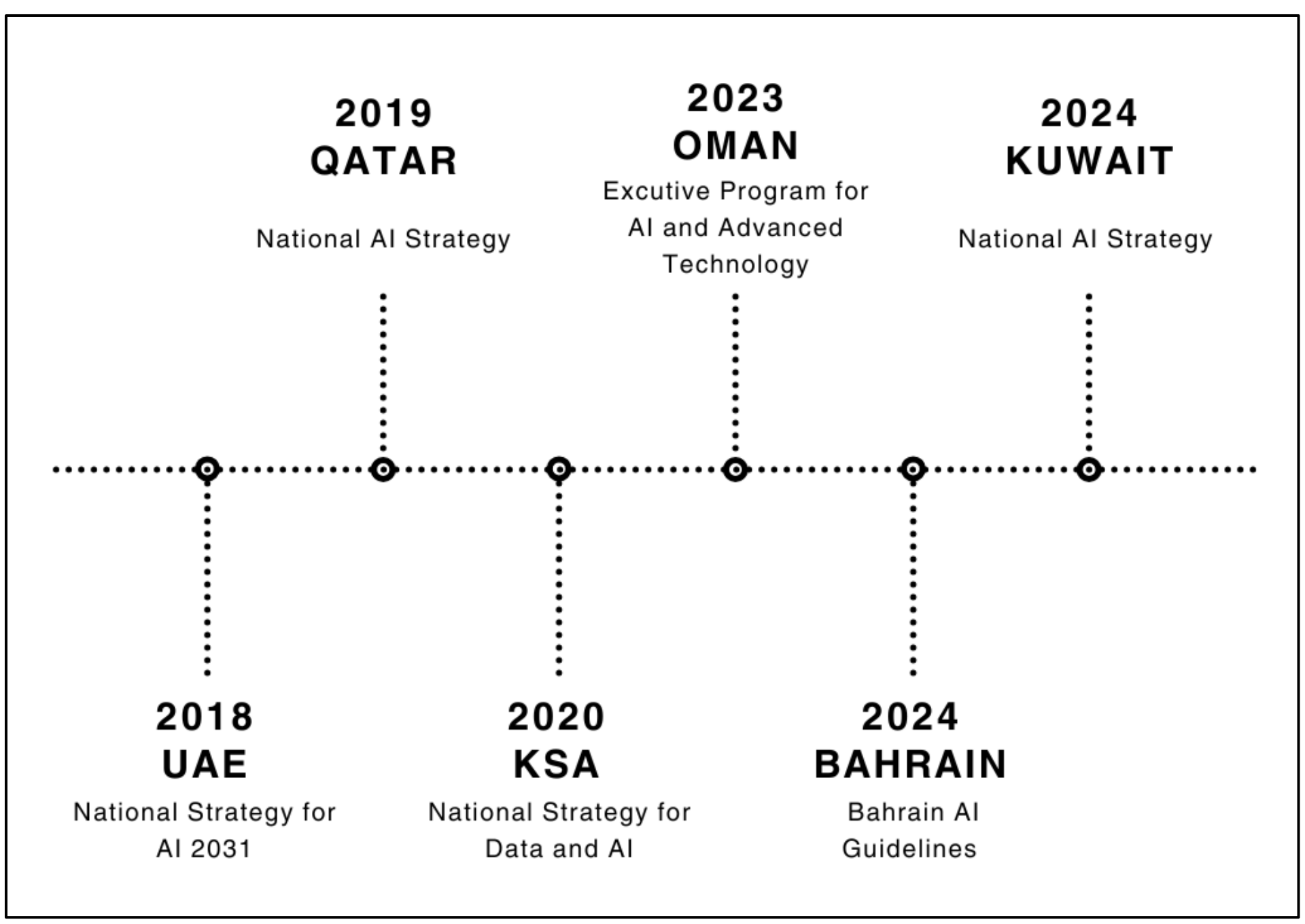

Figure 2. Timeline of NAS Adoption in the GCC

Moreover, the GCC's advanced ICT infrastructure provides a strong foundation for AI initiatives. According to the 2024 ICT Development Index (IDI) report by the International Telecommunication Union (ITU), a specialized agency of the United Nations, all GCC countries score highly, ranging from 91.7 to 100, indicating significant progress in ICT development. The average IDI score among the 170 economies included in the report is 74.8, with the lowest score at 21.3 and the highest at 100. Notably, the GCC countries surpass the average for high-income economies (90.1), reflecting their leadership in global ICT development. Individual country scores further illustrate this trend: Kuwait (100), Qatar (97.8), Bahrain (97.5), the UAE (97.5), Saudi Arabia (95.7), and Oman (91.7). These robust ICT foundations underpin the region's rapid rollout of AI-driven public services.

Institutional developments reinforce this focus: Saudi Arabia's Saudi Authority for Data and Artificial Intelligence (SDAIA) unifies AI and data strategies, while the UAE's first world Ministry of Artificial Intelligence and Mohammed Bin Zayed University of AI (MBZUAI) reflect a concerted push toward national AI expertise. Indeed, GCC governments have enacted top-down mandates that expedite AI implementation, providing ministries and agencies with clear instructions, ample funding, and frequently updated governance frameworks. Despite this rapid pace, the literature notes that implementation-focused studies remain scarce (Aboramadan et al., 2024; Hendawy & Kumar, 2024), indicating a gap between policy intent and empirical validation.

### 2.4.2 Why the GCC Context Is Distinctive

Several factors set the GCC apart from other regions, underscoring the need for a tailored AI Adoption Index. First, centralized governance and strong policy mandates drive much of the GCC's technology adoption. Second, the GCC's resource-rich economies enable large-scale AI deployments, including significant investments in 5G, cloud computing, and specialized research centers. Third, an emphasis on top-down directives fosters rapid pilot projects but can require careful oversight to ensure alignment with ethical and regulatory standards, especially given the relatively nascent state of binding AI legislation across the region.

Further, cultural and demographic influences introduce complexities that generic indices rarely capture. High expatriate populations, diverse cultural norms, and varying levels of digital literacy shape AI workforce development and public engagement with AI systems. Some literature points to the growing ethical and social considerations in GCC AI governance, such as the role of Islamic ethics (Rabbani et al., 2022; Rahayu et al., 2023), yet these dimensions often receive less attention compared to economic imperatives (Crupi & Schilirò, 2023). One illustrative example is Qatar, whose NAS explicitly states that "the framework to be developed must be consistent with Qatari social, cultural, and religious norms", underscoring the importance of aligning AI initiatives with local values and traditions. As a result, global e-government indices and generic AI readiness tools fail to address the policy, economic, and cultural dynamics at play in the GCC, reinforcing the case for a region-specific AI Adoption Index.

### 2.4.3 Opportunities and Challenges

Although the GCC has made considerable progress in formulating comprehensive NASs and ethical guidelines, legally binding AI-specific legislation remains limited across member states and is not yet encompassed in a single, comprehensive act. In many cases, AI-related provisions are integrated into existing data protection and cybercrime laws, focusing on data privacy, algorithmic transparency, security, and liability. Table 1 provides an overview of these provisions, revealing that while each GCC country has taken steps to address AI concerns, these measures remain largely piecemeal rather than consolidated under a dedicated legislative framework. Notably, only the UAE and Saudi Arabia have personal data protection laws nearing the comprehensiveness of the EU's General Data Protection Regulation (GDPR), whereas the other GCC states have yet to establish equally robust PDPL regulations.

Table 1. Overview of AI-related Provisions in Existing GCC Laws

| Country | Law/Regulation |
| --- | --- |
| UAE | UAE Federal Personal Data Protection Law (2021) |
| | Cybercrime Law (Federal Decree-Law No. 34 of 2021) |
| KSA | KSA Personal Data Protection Law (2023) |
| Oman | Royal Decree No. 6/2022 promulgating the Personal Data Protection Law (2022) |
| Qatar | Law No. 13 of 2016 on the Protection of Personal Data |
| | National Cyber Security Agency (NCSA) Guidelines for Secure Adopting and Usage of AI |
| Bahrain | Personal Data Protection Law (PDPL), Law No. 30 of 2019 |
| Kuwait | Data Privacy Protection Regulation No. 26 of 2024 |

Building on these foundational provisions, the GCC has also witnessed a surge in AI-related policy documents, including NASs and AI governance frameworks, since 2018. Notably, all six GCC states have published NASs, with the UAE and Saudi Arabia leading in the volume of additional AI governance documents. Figure 3 illustrates this growing focus, showing how the number of such governance instruments spiked notably in 2024. Despite this top-down momentum, few of these documents are legally binding; rather, they function as strategic or advisory guidelines, underscoring the region's preference for incremental policy evolution. In turn, the lack of a fully enforceable legal framework can reduce incentives for diverse stakeholders, such as private firms, non-governmental organizations, or individuals, to engage actively in AI initiatives, given uncertainty around compliance, accountability, and liability. As a result, the overall impact of these documents on real-world AI deployments and outcomes remains difficult to ascertain.

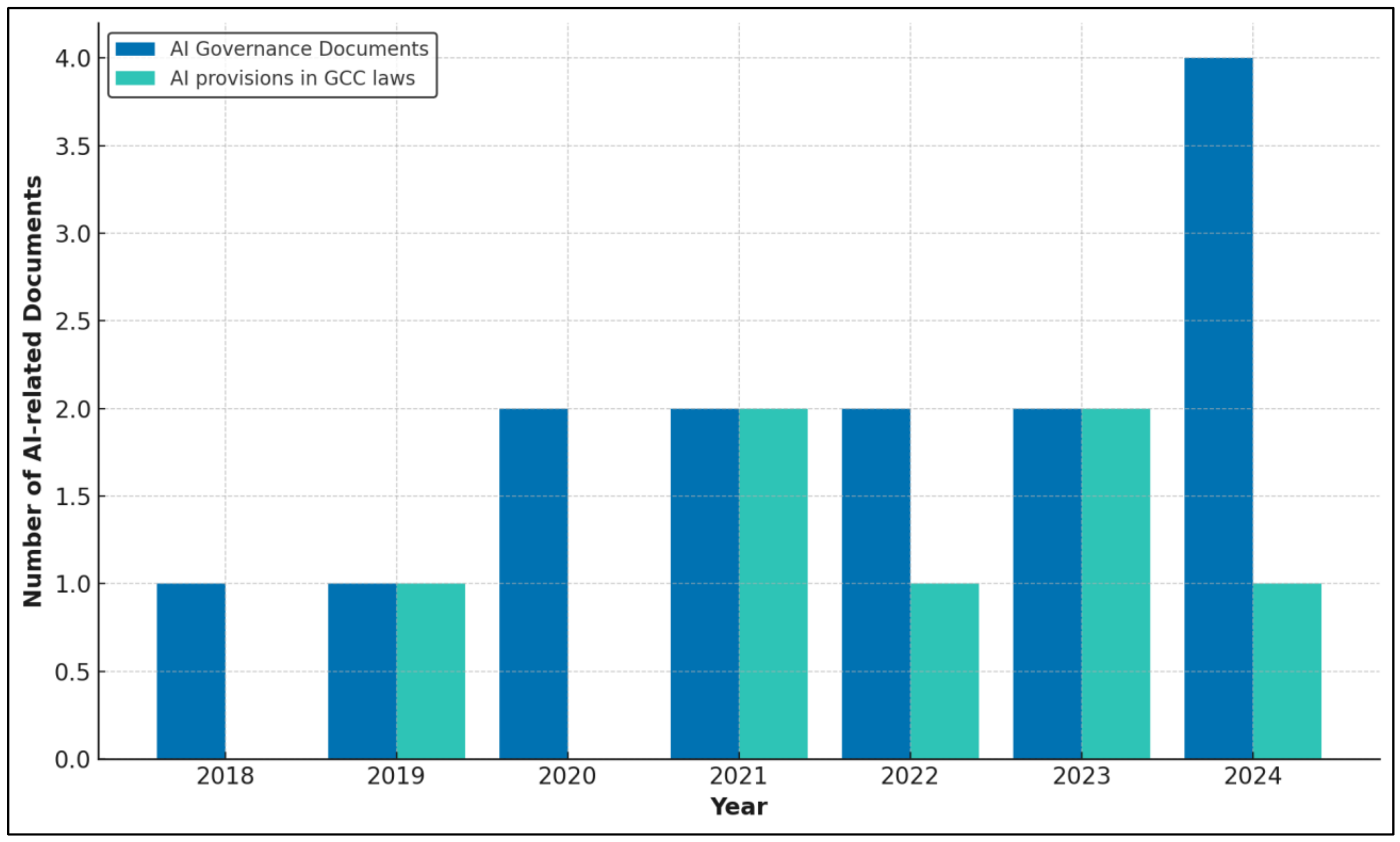

Figure 3. Growth in AI Related Documents in the GCC

In parallel with these legislative and regulatory developments, GCC countries have also advanced Arabic-focused generative AI solutions that illustrate the region's drive to localize emerging technologies. Adapting artificial intelligence to a nation's unique context typically involves a range of activities, policy-making, infrastructure development, and data localization. The UAE's release of Jais (first Arabic large language model built through a collaboration between MBZUAI, G42, Inception, and Cerebras) has been a high-profile example of a Gulf country pushing into Arabic-focused generative AI. While Saudi Arabia developed Noor, a large Arabic language model from King Abdullah University of Science and Technology (KAUST). Noor is a large-scale Arabic language model developed by researchers at KAUST, along with other partners. Launched in 2022, it was one of the largest Arabic LLMs at that time (with around 10 billion parameters). These nation-specific LLM initiatives also highlight the urgency of aligning cutting-edge AI innovations with ethical principles, a theme consistently emphasized in GCC National AI Strategies.

All six GCC NASs emphasize the importance of ethical principles in AI to ensure societal benefits. Table 2 provides an overview of each principle's Term Frequency–Inverse Document Frequency (TF-IDF), including the document frequency (df), the number of NASs referencing it (out of six), and a corresponding binary TF-IDF calculation. Details about the underlying methodology, such as assigning TF = 1 when a principle is mentioned and defining IDF as LN(6 / df), can be found in Section 3.2. By comparing average TF-IDF values, we can identify widely recognized principles, such as Privacy & Data Protection, as well as those that receive comparatively less attention, such as Benefit to Humanity. While AI is acknowledged as a tool for human empowerment in these strategies, economic growth and innovation often take precedence over the explicit prioritization of maintaining human agency and control. This suggests a potential area for further development in GCC AI governance to ensure that AI genuinely serves and benefits the population. Moreover, although ethical principles like Privacy & Data Protection are emphasized, achieving meaningful compliance requires robust data protection laws across GCC states. Without legally binding mechanisms to enforce these ethical commitments, they risk functioning more as aspirational statements than as practical obligations, a phenomenon sometimes described as "ethicswashing" (Schultz et al., 2024).

Table 2: Frequency of Ethical Principles in GCC NASs Documents

| Ethical Principle | Doc. Freq. (df) | IDF = LN (6 / df) | Countries Mentioned | Avg. TF-IDF |
|---|---|---|---|---|
| Privacy & Data Protection | 6 | 0.00 | KSA, UAE, Kuwait, Bahrain, Qatar, Oman | 0.00 |
| Accountability | 5 | 0.18 | KSA, UAE, Kuwait, Bahrain, Qatar | 0.15 (approx.) |
| Transparency | 5 | 0.18 | KSA, UAE, Kuwait, Bahrain, Oman | 0.15 (approx.) |
| Fairness & Non-discrimination | 4 | 0.41 | KSA, UAE, Bahrain, Oman | 0.35 (approx.) |
| Human Oversight & Control | 3 | 0.69 | UAE, Bahrain, Oman | 0.58 (approx.) |
| Robustness & Safety | 3 | 0.69 | UAE, Kuwait, Oman | 0.58 (approx.) |
| Sustainability & Environmental Friendliness | 2 | 1.10 | UAE, Bahrain | 0.90 (approx.) |
| Human-cantered Values | 2 | 1.10 | UAE, Oman | 0.90 (approx.) |
| Collaboration & Inclusivity | 1 | 1.79 | KSA | 1.79 (approx.) |
| Benefit to Humanity | 1 | 1.79 | Qatar | 1.79 (approx.) |

Beyond legal and policy frameworks, most NASs emphasize the urgent need for capacity-building investments to ensure sustainable AI development. To elucidate these priorities, a TF-IDF analysis (Manning, 2009; Robertson, 1977) was conducted on all instances where key terms appeared in the NAS texts, whether in titles, informational statements, or normative directives. This analysis confirms that capability development and workforce readiness rank among the most frequently mentioned concepts. Figure 4 visualizes which AI capabilities, such as "AI ethics and governance", "cybersecurity", and "NLP", are emphasized across the region. While frequency alone cannot capture each country's unique context, these findings reveal overarching themes that policymakers perceive as vital for successful AI adoption. More details on the TF-IDF values for Figure 4 are provided in the online codebook (see Section 3.8).

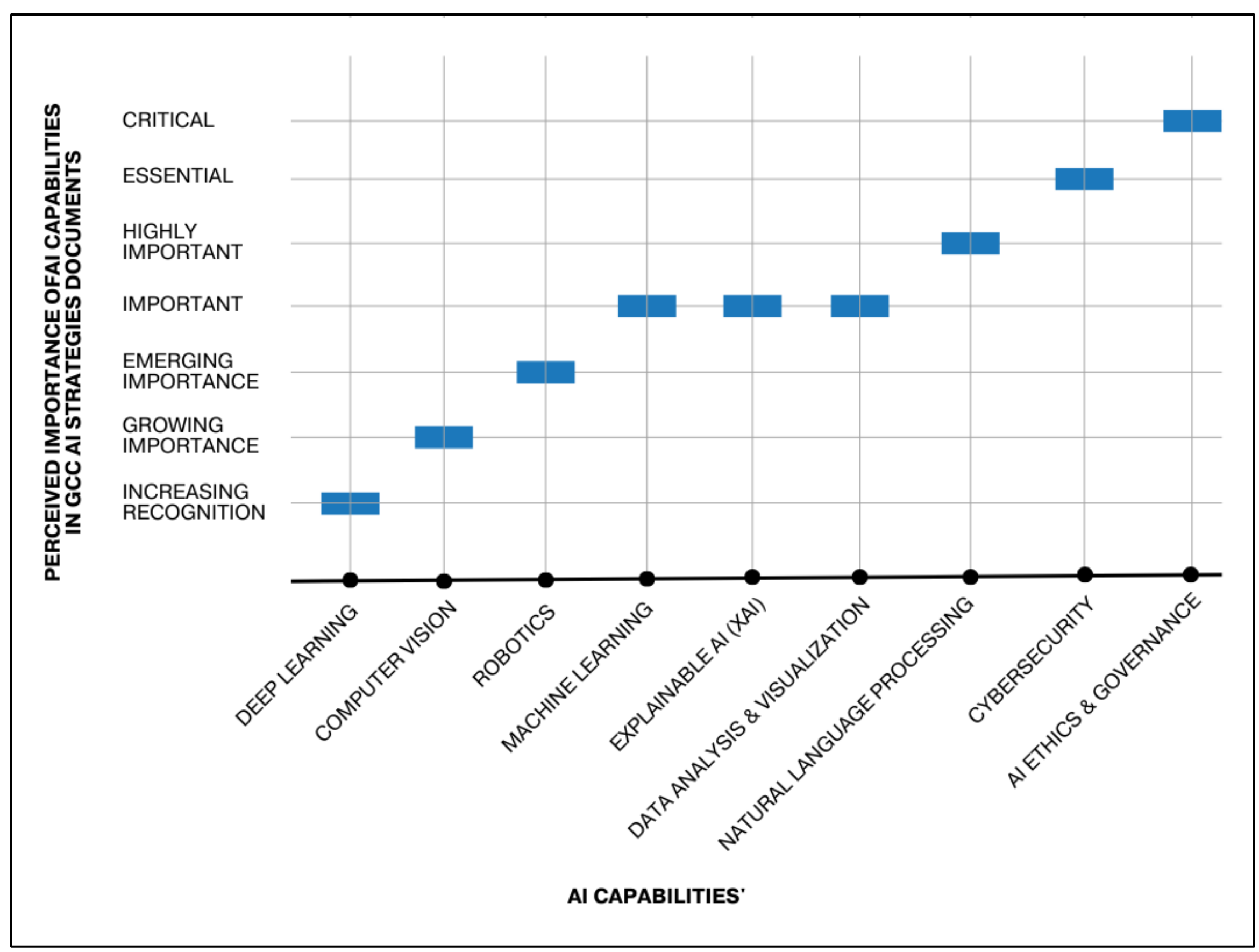

Figure 4. Bar Chart showing the Perceived Importance of different AI Capabilities in GCC NASs Documents

However, no standardized tool currently exists to measure how these evolving policy directives, legislative provisions, and capability-building efforts translate into tangible improvements, such as enhanced citizen satisfaction or greater service efficiency. These circumstances highlight the need for a dedicated index capable of tracking both technical and policy milestones in GCC AI adoption. By systematically gauging whether strategic priorities and NAS objectives materialize in practical outcomes, researchers and decision-makers can identify gaps, target resources, and refine strategies for sustained, AI-driven transformation in the region.

### 2.5 Gaps and Rationale for a GCC-Specific AI Adoption Index

Despite the availability of e-government and digital maturity indices, several gaps emerge when these tools are applied to the GCC. For example, the World Economic Forum's Global Competitiveness Report (2020) demonstrates how high-level competitiveness metrics often overlook localized governance structures and context-specific policy mandates. Traditional metrics also rarely consider top-down governance, the approach through which AI directives are mandated by central authorities rather than evolving from grassroots or market-driven initiatives. As a result, conventional indices do not adequately measure how policy mandates shape the rapid rollout, funding mechanisms, or accountability structures for AI projects. Meanwhile, the UN E-Government Survey (2024) and the World Bank's GovTech Maturity Index (2022) do track progress in digital service provision, yet they rarely account for the high-resource investments and rapid AI deployments characteristic of GCC nations. Moreover, OECD (2024) points out that AI governance models must address ethical, regulatory, and workforce factors, dimensions often absent or insufficiently weighted in generic digital indices.

Further complicating matters is the resource-intense nature of GCC economies, which allows for large-scale AI implementations far exceeding the incremental technology adoption assumed by many existing frameworks. Standard indices often fail to capture the speed and scale of these deployments, overlooking how well-funded infrastructure and robust government backing can accelerate AI priorities and outcomes. Likewise, broader digital benchmarks typically focus on basic connectivity and online services, offering limited insight into advanced AI capabilities such as data analytics, predictive modeling, and algorithmic decision-making. This oversight neglects the fact that GCC nations frequently pursue cutting-edge solutions, an approach that invariably raises new ethical, regulatory, and capacity-building questions. Finally, cultural and demographic factors, including high expatriate populations, religious considerations, and multilingual service needs, add a further layer of complexity. Existing studies also highlight the GCC's limited representation in global AI ethics debates, raising concerns that regional nuances may be underexplored in mainstream AI governance frameworks (Roche et al., 2023).

Given these gaps, the unique context of the GCC underscores the need for a specialized AI Adoption Index. By tailoring the index to the region's governance style, economic capabilities, infrastructural baselines, regulatory evolution, and cultural realities, researchers and policymakers can more accurately gauge AI maturity and strategically guide future public-sector transformations. Indeed, these limitations underscore why a GCC-specific tool is critical. Drawing on insights from the Global Economic Diversification Index (Prasad et al., 2025), particularly its emphasis on linking technology investments to non-oil revenue growth, would illuminate how AI initiatives align with broader diversification agendas. Such a specialized index can, therefore, provide actionable insights for policymakers, helping them identify both technical readiness shortfalls and governance vulnerabilities. By addressing the unique drivers, barriers, and cultural contexts shaping AI adoption in the GCC, this framework promises to offer a robust foundation for sustained, AI-driven public-sector innovation in the region.

## 3. Methodology

To ensure our framework captured both formal policy objectives and real-world practice, we adopted a two-stage approach. First, we conducted a top-down content analysis of official policy documents (the six GCC NASs) and relevant literature, extracting recurring themes in AI governance. Second, we complemented these findings with a bottom-up quantitative study, a survey of 203 mid- and senior-level government employees. This integrated methodology allows us to test whether the policy-derived dimensions (Technical Infrastructure, Organizational Readiness, and Governance Environment) actually predict tangible AI adoption outcomes within the region's public-sector contexts. This study adopted a multi-stage approach that is both theory- and data-driven. First, we conducted a document analysis of relevant literature and six GCC NASs, extracting key theoretical constructs related to AI adoption (Technical Infrastructure, Organizational Readiness, and Governance Environment). Second, we employed quantitative, data-driven methods, namely, a survey of 203 government employees and advanced statistical techniques (K-Means clustering, PCA, and Partial Least Squares Structural Equation Modeling), to empirically test and validate these constructs. This two-tiered methodology ensures that the final AI Adoption Index reflects both policy-based theory and real-world data (for research design, see Figure 5).

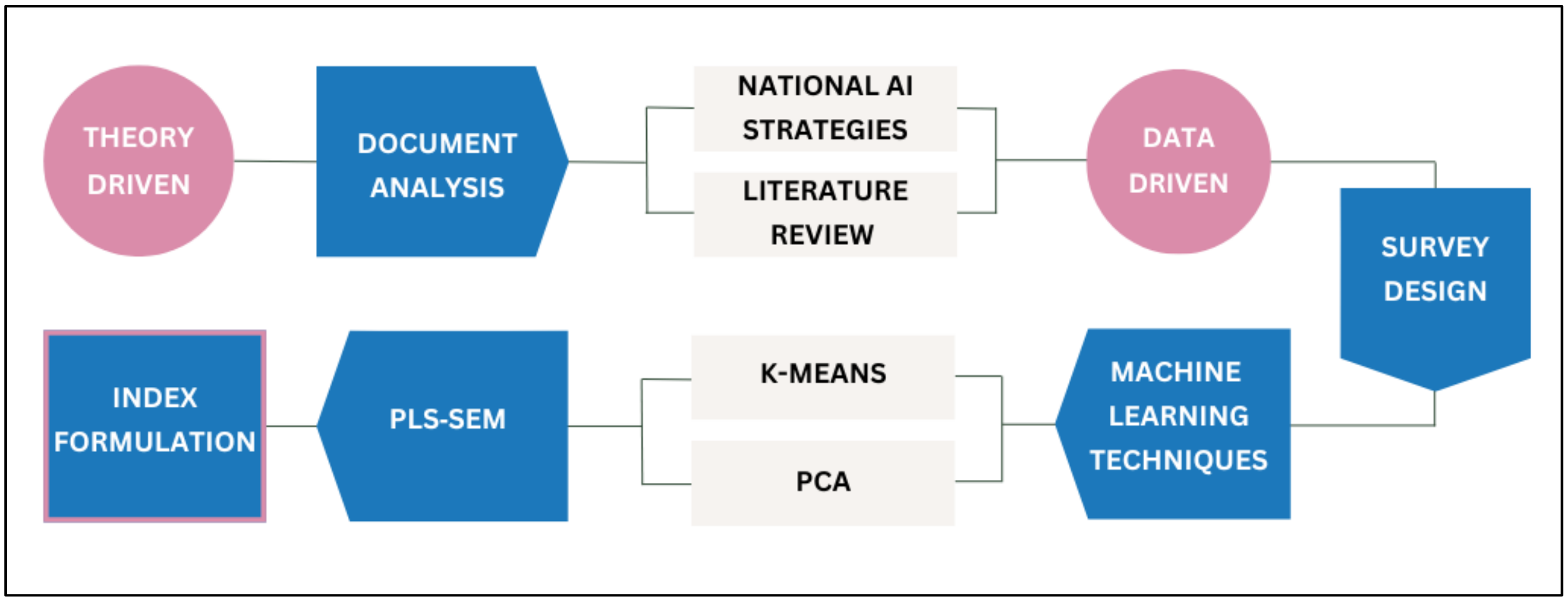

Figure 5. Overview of Research Design

All code, data, and procedural details are available in online repositories to ensure reproducibility and transparency (see Section 3.8).

### 3.1 Theoretical Underpinnings

Building on e-government and AI readiness literature, this study draws on the TAM (Davis, 1989), which emphasizes perceived usefulness and organizational readiness as key drivers of technology adoption. GCC-specific AI policy frameworks further refined these constructs to incorporate infrastructural, policy, and regulatory elements, reflecting regional priorities in governance, data security, and resource investments. By synthesizing both TAM and GCC policy insights, the conceptual foundation addressed how technical infrastructural, organizational readiness, and Governance Environment alignment might influence AI Outcomes.

### 3.2 Policy Document Analysis

Each GCC state, Bahrain, Kuwait, Oman, Qatar, Saudi Arabia, and the UAE, maintains an official NAS document. These six NAS documents were selected as the primary sources for this research, as each country has only one such strategy, eliminating the need for further exclusion. To address potential language barriers, we relied on professional translation services for Arabic texts, followed by back-translation to English to verify accuracy. A multilingual glossary of key AI terms in English and Arabic was developed to maintain terminological consistency.

All NAS documents were converted to machine-readable text using the pdfplumber Python library, with pages or sections rendered as images processed via Adobe's online OCR service and the Tesseract Python library. Approximately 17% of the NAS corpus required OCR, with an estimated error rate below 5%. This preparation ensured that the documents were fully searchable and analyzable for both qualitative and quantitative procedures.

Two researchers independently coded each NAS document using a customized codebook designed to capture capacity-building references. Nine thematic categories relevant to AI governance, ranging from Infrastructure & ICT to Data Governance & Privacy, were defined, along with specific guidelines for flagging text related to training, skill gaps, or organizational readiness. To measure inter-coder consistency, Cohen's kappa (Cohen, 1960) was calculated, yielding a value above 0.79, which indicates substantial reliability (Landis & Koch, 1977).

For codebook validity, a TF–IDF procedure is introduced to quantify the relative importance of specific AI capabilities (for example, Deep Learning, Robotics, Cybersecurity) in capacity-building contexts. The TF–IDF computation is expressed in this equation:

$$TF - IDF(t, d) = TF(t, d) \times IDF(t)$$

Where $TF(t, d)$ measures the frequency of term $t$ in document $d$, and $IDF(t)$ indicates how unique $t$ is across the corpus. This mixed-method approach, encompassing both qualitative coding and TF–IDF analysis, ensures consistent classification of thematic content and allows for seamless integration with subsequent survey findings. Full details on the coding categories, TF–IDF calculations, and illustrative examples appear in online dataset (see section 3.8).

Finally, to contextualize the survey, the thematic coding and TF–IDF insights from these official AI strategy and policy documents informed the focus on Technical Infrastructure, Organizational Readiness, and the Governance Environment. These focal areas aligned closely with recurring capacity-building priorities in the NAS corpus and thus provided a robust framework for shaping the survey instrument.

### 3.3 Data Cleaning and Pre-Processing

Before proceeding with statistical analyses, data cleaning was performed on the survey responses. No items were missing from the final dataset; therefore, no imputation was required. Outliers were examined via boxplots; no extreme outliers were removed, as they appeared to represent genuine variations rather than data errors. All Likert-scale items (Q2–Q12) were standardized (z-scores) prior to clustering to ensure that variables with larger numerical ranges did not unduly influence the distance measures used in K-Means.

### 3.4 Survey Design and Sampling

To develop the survey instrument, the research team first adapted items from validated TAM scales (Davis, 1989) and aligned them with GCC-specific AI policy frameworks. Additional validation references drawn from prior e-government and AI readiness studies guided the wording of each item, ensuring content and construct validity. The final questionnaire included eleven Likert-scale items (Q2–Q12) focused on perceived AI usefulness, organizational readiness, policy alignment, and anticipated outcomes, plus an additional item (Q1) capturing demographic or categorical information (country). Before wide deployment, a pilot test was conducted with ten mid-level government professionals from two GCC ministries to evaluate question clarity, relevance, and use of technical terminology. Feedback from this pilot resulted in minor revisions to item phrasing and response formats.

The target population comprised mid-to-senior employees in AI/IT roles across government agencies and professional associations in the GCC. These individuals were deemed most directly involved in AI implementation and policy decisions; thus, their insights would be particularly relevant to assessing AI adoption levels. A purposive sampling approach was employed, drawing on official government directories and professional association membership lists. While this strategy ensured domain expertise, it also risked overrepresenting respondents with heightened technology awareness.

Data collection occurred between September 2024 and February 2025. An online survey link was disseminated via email invitations sent to over 400 potential participants, ultimately yielding 203 valid and complete responses (50.7% response rate). Relying on an online distribution method minimized logistical barriers across multiple GCC states and likely enhanced participant candor by reducing Hawthorne effects (Adair, 1984), yet it also restricted opportunities for in-person follow-ups. All participants provided informed consent in accordance with institutional ethical guidelines.

To detect any systematic differences between respondents and non-respondents, a non-response bias check compared early respondents (those completing the survey in the first three weeks) with late respondents (those in the final three weeks) on key demographic variables such as years of experience and job level. No statistically significant differences emerged ($p > 0.05$), suggesting minimal non-response bias. Overall, this methodologically rigorous process, grounded in TAM concepts, GCC policy themes, and pilot feedback, laid a strong foundation for assessing AI adoption levels in the region.

## 3.5 Machine-Learning Techniques

### 3.5.1 K-Means Clustering

K-Means clustering was applied to the standardized scores of the primary Likert-scale items (Q2–Q12) to identify underlying groupings (clusters) of respondents. Prior to clustering, each variable was standardized to z-scores to ensure all items had comparable scales. Multiple values of $k$ (ranging from 2 to 6) were tested, and the elbow method was used to select the optimal number of clusters. Specifically, the sum of squared distances between each data point and its assigned cluster centroid was plotted against increasing values of $k$; the point at which the rate of decrease visibly "leveled off" indicated a suitable choice for $k$ (Ketchen & Shook, 1996; Steinley & Brusco, 2011).

Once the optimal $k$ was determined, K-Means was performed iteratively until convergence, producing cohesive groups of respondents. Although no findings are reported in this section, it is standard practice to label the resulting clusters based on emergent patterns in the data and to align these labels with relevant frameworks in technology adoption and organizational readiness literature. This labeling process typically involves examining average scores on the clustering variables and comparing them to conceptual expectations drawn from existing theory. By adhering to these procedures, the study ensures that each cluster can be meaningfully interpreted at later stages of analysis.

### 3.5.2 Principal Component Analysis (PCA)

Following the clustering procedures, Principal Component Analysis (PCA) was employed to determine whether the eleven Likert-scale items (Q2–Q12) consolidated into one or more latent factors. A combination of eigenvalues > 1, the scree plot, and parallel analysis guided the decision on how many components to retain, and an initial Varimax rotation was applied to identify clear factor structures. The study also tested forced multi-factor solutions to assess the unidimensional or multi-dimensional nature of AI adoption constructs.

In handling cross-loadings, defined as items loading above 0.30–0.40 on multiple factors, each item was closely reviewed for conceptual overlap. When persistent cross-loadings could not be explained by theory, such items were either retained in the factor most consistent with prior literature or removed to enhance factor clarity. This approach ensured that each principal component was both theoretically interpretable and distinct, laying a robust foundation for subsequent analyses of AI adoption constructs.

## 3.6 Partial Least Squares Structural Equation Modeling (PLS-SEM)

To assess the relationships among the key latent variables in this study, Partial Least Squares Structural Equation Modeling (PLS-SEM) was employed using SmartPLS 4. This choice was guided by the exploratory nature of the research, the complex interplay of multiple latent variables, and a sample size smaller than is typically required by covariance-based SEM (Hair et al., 2022; Sarstedt et al., 2022). In addition, PLS-SEM is robust to non-normal data and offers flexibility in modeling both formative and reflective constructs. These characteristics make PLS-SEM well-suited for developing an AI Adoption Formula that links technical infrastructure, organizational readiness, and governance environment factors to overall AI outcomes.

### 3.6.1 Rationale and Hypotheses

In alignment with TAM's focus on perceived usefulness and readiness, and incorporating GCC policy themes on infrastructure and governance, we define and operationalize these constructs as follows:

- **Technical Infrastructure (TI):** ICT and data foundations (such as networks, cloud, compute, platforms, datasets) that enable AI deployment.
- **Organizational Readiness (OR):** Internal skills, leadership support, and processes that allow public agencies to adopt and manage AI.
- **Governance Environment (GE):** External policies, regulations, and ethical oversight shaping responsible and effective AI use.
- **AI Outcomes (AIO):** Perceived benefits from AI adoption, including service efficiency, citizen satisfaction, and performance.

Based on these definitions, the following hypotheses were formulated:

- **H1:** Technical Infrastructure → AI Outcomes
- **H2:** Organizational Readiness → AI Outcomes
- **H3:** Governance Environment → AI Outcomes

By modeling these constructs, the approach aims to clarify how resources, culture, and policy frameworks collectively influence AI adoption and outcomes in GCC public-sector contexts.

### 3.6.2 Model Specification and Evaluation

Within SmartPLS 4, latent constructs were defined for Technical Infrastructure, Organizational Readiness, Governance Environment, and AI Outcomes, employing a path weighting scheme for analysis. Below are the key steps and findings from the measurement model checks, structural model estimates, and potential index derivation.

- **Reflective vs. Formative Constructs**

Two latent variables, Technical Infrastructure and Organizational Readiness, were modeled reflectively, as the observed items were treated as manifestations of their underlying constructs. Conversely, Governance Environment was modeled as a formative construct, reflecting the idea that policies, regulations, and guidelines collectively contribute to its conceptual domain rather than merely reflect it. This distinction necessitated different approaches for validity and reliability assessments within the measurement model.

- **Common Method Variance**

Because data were collected via self-report surveys, common method variance was tested using Harman's single-factor test. The first factor did not account for the majority of variance, suggesting that common method bias was not a major concern. A full collinearity test in SmartPLS also found no significant issues indicative of method bias.

**A. Measurement Model Checks.**

Following Kline's (2015) guidelines, we examined Cronbach's alpha, Composite Reliability, and Average Variance Extracted (AVE) for reflective constructs. For formative constructs, the focus shifted to outer weights and loadings, evaluating their significance to confirm construct validity.

- **Discriminant Validity:**

Assessed using the Fornell–Larcker criterion (Fornell & Larcker, 1981), ensuring each construct's AVE exceeded the squared correlation with other constructs. Additionally, the Heterotrait–Monotrait (HTMT) ratio (Henseler et al., 2015) provided a more robust check for discriminant validity.

- **Model Fit Indices in PLS-SEM:**

Although model fit in PLS-SEM is often debated, we report the Standardized Root Mean Square Residual (SRMR) and the Normed Fit Index (NFI) as supplementary indicators of model adequacy.

**B. Structural Model Estimates**

After confirming the measurement model's adequacy, the structural model was evaluated by examining path coefficients (β), the coefficient of determination ($R^2$), and bootstrapped significance tests with 5,000 resamples. This procedure determined the strength and significance of relationships among the latent constructs, directly addressing H1, H2, and H3.

**C. Index Derivation**

In instances where Principal Component Analysis (PCA) indicated unidimensionality for certain constructs, factor loadings or path weights could be integrated into a single "AI Adoption Index". Such an index serves as a practical tool for policymakers, facilitating the measurement of AI adoption levels and highlighting priority areas for intervention or support across different GCC entities.

### 3.7 Limitations

While the online survey facilitated data collection across multiple GCC countries by minimizing geographical barriers and travel burdens, it also limited opportunities for face-to-face validation. Relying on an online format may introduce selection bias, as the purposive sampling of AI/IT professionals could overrepresent individuals with higher technology awareness, potentially inflating perceived readiness and outcomes. This targeted recruitment strategy, while beneficial for eliciting domain-specific insights, may not capture the perspectives of less technologically inclined stakeholders.

Furthermore, forcing single-factor or multi-factor solutions in Principal Component Analysis (PCA) could mask important sub-dimensions such as ethics, data privacy, or local cultural considerations. The choice of factor structure remains contingent on item design and correlation patterns; thus, any unidimensional or multi-dimensional model may overlook nuanced aspects of AI adoption.

To triangulate findings and deepen contextual understanding, future research could employ mixed methods, including interviews, focus groups, or expanded survey instruments. Such approaches would allow for a more granular exploration of diverse viewpoints and additional thematic areas, ultimately offering a more holistic portrayal of AI adoption within the GCC.

### 3.8 Data Repository Access

To ensure transparency and reproducibility, the following materials are provided as appendices in separate files hosted in the Harvard Dataverse repository:

- **Document Analysis Codebook.** A customized codebook developed to analyse capacity-building references in the six GCC NASs, defining nine thematic categories relevant to AI governance. Available at: https://doi.org/10.7910/DVN/3LAOVI
- **Anonymized Survey Responses.** The raw, individual-level responses to Q1–Q12, with all personal identifiers removed to preserve participant privacy. Available at: https://doi.org/10.7910/DVN/60PTXQ
- **Final SmartPLS Report**. Contains an Excel file of the cleaned dataset and key PLS-SEM outputs used in this study. Available at: https://doi.org/10.7910/DVN/OM42YL

Each repository includes brief readme notes that describe the purpose of each file and guidelines for use. Researchers interested in replicating or extending the findings can use these materials to follow our analytic approach and validate the results reported in this paper.

## 4. Results

In line with this perspective, the following subsections present descriptive data, K-Means clustering outcomes, Principal Component Analysis (PCA) findings, and the PLS-SEM model, concluding with an AI Adoption Index equation. All quantitative analyses were conducted via IBM SPSS (for K-Means, PCA) and SmartPLS 4 (for SEM).

### 4.1 Survey Descriptive Statistics

A total of 203 valid responses were collected from mid- and senior-level government employees in AI or IT roles across GCC ministries and agencies. Table 3 summarizes respondents' country distribution, showing representation from all six GCC countries

Table 3. Survey Sample

| Country | Count (n) | Percent (%) |
|---|---|---|
| Saudi Arabia | 34 | 16.7% |
| UAE | 35 | 17.2% |
| Qatar | 33 | 16.3% |
| Bahrain | 34 | 16.7% |
| Kuwait | 35 | 17.2% |
| Oman | 32 | 15.8% |
| Total | 203 | 100.0% |

As shown in Table 4, the average scores for the eleven Likert-scale items (Q2–Q12) ranged from 3.18 (Q9) to 3.68 (Q11), with standard deviations typically falling between 0.89 and 1.06, indicating moderate variance in perceived AI adoption. Additionally, no significant multicollinearity concerns were observed based on initial diagnostics.

Table 4. Descriptive Statistics and Inter-Item Correlations

| Constructs | Item | Mean | Std. Dev. | Min | Max |
|---|---|---|---|---|---|
| Technical Infrastructure (TI) | Q2 | 3.56 | 1.005 | 1 | 5 |
| | Q3 | 3.61 | 1.035 | 2 | 5 |
| | Q4 | 3.58 | 1.008 | 1 | 5 |
| Organizational Readiness (OR) | Q5 | 3.46 | 1.035 | 2 | 5 |
| | Q6 | 3.39 | .908 | 2 | 5 |
| | Q7 | 3.50 | 1.045 | 1 | 5 |
| Governance Environment (GE) | Q8 | 3.25 | 1.049 | 2 | 5 |
| | Q9 | 3.18 | 1.058 | 1 | 5 |
| AI Outcomes (AIO) | Q10 | 3.66 | .889 | 2 | 5 |
| | Q11 | 3.68 | .891 | 2 | 5 |
| | Q12 | 3.65 | .886 | 2 | 5 |

A closer look at the individual scales reveals strong, positive correlations among items intended to measure the same construct. For the Technical Infrastructure scale, inter-item correlations ranged from 0.835 to 0.905 ($p < .001$), indicating a tightly linked set of perceptions regarding technological infrastructure. Organizational Readiness items showed similarly high correlations (0.850–0.941, $p < .001$), suggesting the items capture closely related aspects of internal organizational readiness. The Governance Environment scale (two items) exhibited a correlation of 0.882 ($p < .001$), reflecting a cohesive underlying policy domain. Finally, AI Outcomes items were moderately to strongly intercorrelated (0.784–0.819, $p < .001$), pointing to consistent views on the benefits and impacts of AI initiatives. These findings support the unidimensional nature of each scale and confirm that each set of items appropriately reflects a single underlying construct.

To confirm each set of items measured a cohesive underlying construct, Cronbach's alpha was computed for the four scales. The resulting values indicated excellent internal consistency: Technical Infrastructure ($\alpha = 0.949$), Organizational Readiness ($\alpha = 0.959$), Governance Environment ($\alpha = 0.937$), and AI Outcomes ($\alpha = 0.924$). Following this reliability check, all Likert-scale items were standardized (z-scores) to ensure variables with different scales did not unduly influence distance calculations in the K-Means procedure. This step was supported by descriptive checks showing variability across item means and standard deviations.

## 4.2 K-Means Clustering

### 4.2.1 Selection of Cluster Count

To explore empirical groupings among respondents, K-Means clustering was conducted using the standardized Likert-scale items (Q2–Q12). The procedure was run six times ($k = 2\ to\ 6$), aligned with the six GCC states. As illustrated in Figure 6, the most pronounced "elbow" in the within-group sum of squares emerged at $k = 2$, suggesting a two-cluster solution offered the best balance between explanatory power and parsimony.

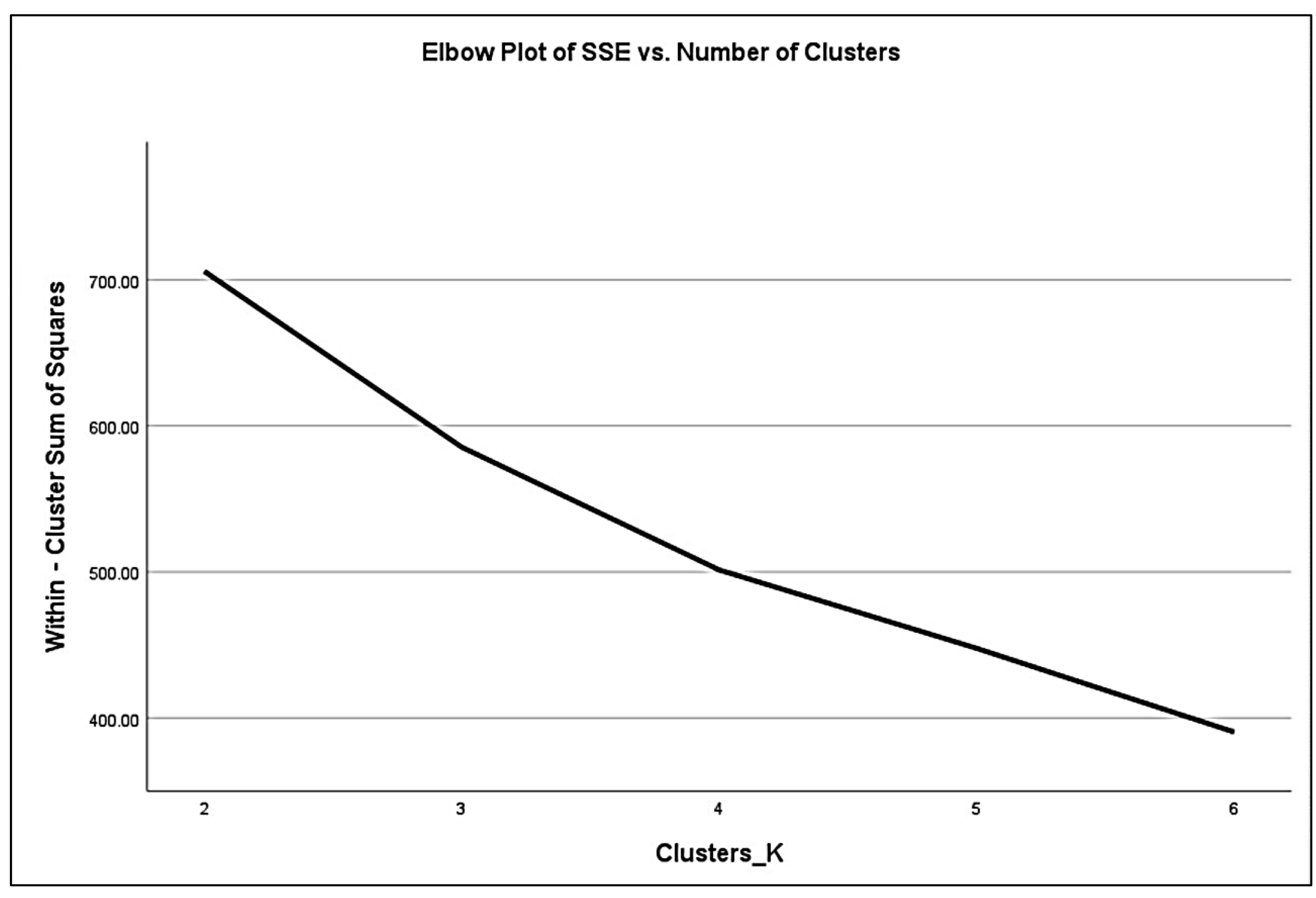

Figure 6. Elbow Plot for K-Means Clustering (k=2–6)

### 4.2.2 Cluster Profiles

In the 2-cluster solution, the first cluster, labeled "High AI Adoption", primarily comprised respondents from UAE and KSA, who exhibited strong agreement (positive z-scores) across technical infrastructure, organizational readiness, and policy support items. By contrast, the second cluster "Moderate Adoption", included other GCC states, reflecting only partial endorsement of these adoption factors. The High AI Adoption group was smaller in size (Two countries), but exhibited consistently higher standardized scores, whereas the Moderate Adoption group encompassed the majority of respondents with mid-range scores.

To verify this demographic distinction, the Q1 (country) variable was recoded in SPSS so that "UAE" and "KSA" formed one category and "other GCC states" formed a second category. As illustrated in Table 5 and 6, a crosstab of the recoded variable with cluster membership indicated that Cluster 1 (High AI Adoption) contained 69 respondents (UAE & KSA) and Cluster 2 (Moderate Adoption) contained 134 respondents (other GCC states).

Table 5. Number of Cases in Each Cluster

| Cluster | Number of cases |
|---|---|
| 1 (High AI Adoption) | 69 |
| 2 (Moderate Adoption) | 134 |
| Valid | 203 |
| Missing | 0 |

Table 6. Final Cluster Centers

| | Cluster 1 | Cluster 2 |
|---|---|---|
| (TI) (Q2) | 5 | 3 |
| (TI) (Q3) | 5 | 3 |
| (TI) (Q4) | 5 | 3 |
| (OR) (Q5) | 5 | 3 |
| (OR) (Q6) | 4 | 3 |
| (OR) (Q7) | 5 | 3 |
| (GE) (Q8) | 4 | 3 |
| (GE) (Q9) | 4 | 3 |
| (AIO) (Q10) | 5 | 3 |
| (AIO) (Q11) | 5 | 3 |
| (AIO) (12) | 5 | 3 |

These cluster labels clarify how different AI Adoption levels are distributed across GCC respondents, thereby offering a practical means to distinguish between high-adoption countries and those with more moderate adoption. This refined perspective contributes to a deeper understanding of the underlying factors shaping AI implementation in the region's public sector. Overall, this K-Means clustering procedure facilitated the identification of distinct adopter profiles within the GCC sample (moderate vs. high adopters). By comparing these empirical clusters against theoretical expectations drawn from both document analysis and prior AI-adoption frameworks, the study addressed potential heterogeneity among respondents and provided a nuanced understanding of how different adoption levels manifest across government AI/IT professionals in the region.

## 4.3 Principal Component Analysis (PCA)

### 4.3.1 Sampling adequacy and Factorability

Before performing PCA, the suitability of the data was assessed. KMO (Kaiser-Meyer-Olkin) was 0.937, indicating excellent sampling adequacy. Additionally, Bartlett's Test of Sphericity was highly significant ($p < .001$), confirming that the correlation matrix was factorable and appropriate for PCA (see Table 7).

Table 7. KMO and Bartlett's Test

| Measure | Value |
|---|---|
| Kaiser–Meyer–Olkin (KMO) | 0.937 |
| Bartlett's Test of Sphericity | $p < .001$ |

### 4.3.2 Communalities

Communalities (Table 8) showed that all items (covering technical infrastructure, organizational readiness, governance environment, and AI outcomes) had strong extraction values (ranging from about 0.66 up to 0.88), suggesting each item shared substantial variance with the underlying factor(s).

Table 8. Communalities (Extraction)

| Item | Extraction range |
|---|---|
| TI | 0.765–0.816 |
| OR | 0.758–0.878 |
| GE | 0.748–0.764 |
| AIO | 0.658–0.669 |

### 4.3.3 Total Variance Explained and Scree Plot

As shown in Table 9, the PCA with eigenvalues over 1 extracted a single component with an eigenvalue of 8.495, accounting for 77.23% of the total variance. The second potential component had an eigenvalue of only 0.880 (< 1), falling below the standard retention threshold. Consequently, only one factor was retained.

Table 9. Initial eigenvalues

| Component | Total | % of Variance | Cumulative % |
|---|---|---|---|
| 1 | 8.495 | 77.231 | 77.231 |
| 2 | .880 | 8.003 | 85.234 |
| 3 | .476 | 4.326 | 89.560 |
| 4 | .223 | 2.027 | 91.587 |
| 5 | .209 | 1.901 | 93.489 |
| 6 | .183 | 1.659 | 95.148 |

| Component | Total | % of Variance | Cumulative % |
|---|---|---|---|
| 7 | .153 | 1.390 | 96.537 |
| 8 | .132 | 1.203 | 97.740 |
| 9 | .117 | 1.064 | 98.804 |
| 10 | .082 | .741 | 99.545 |
| 11 | .050 | .455 | 100.000 |

The Scree Plot in Figure 7 clearly illustrates a steep drop after the first factor, supporting the one-factor solution.

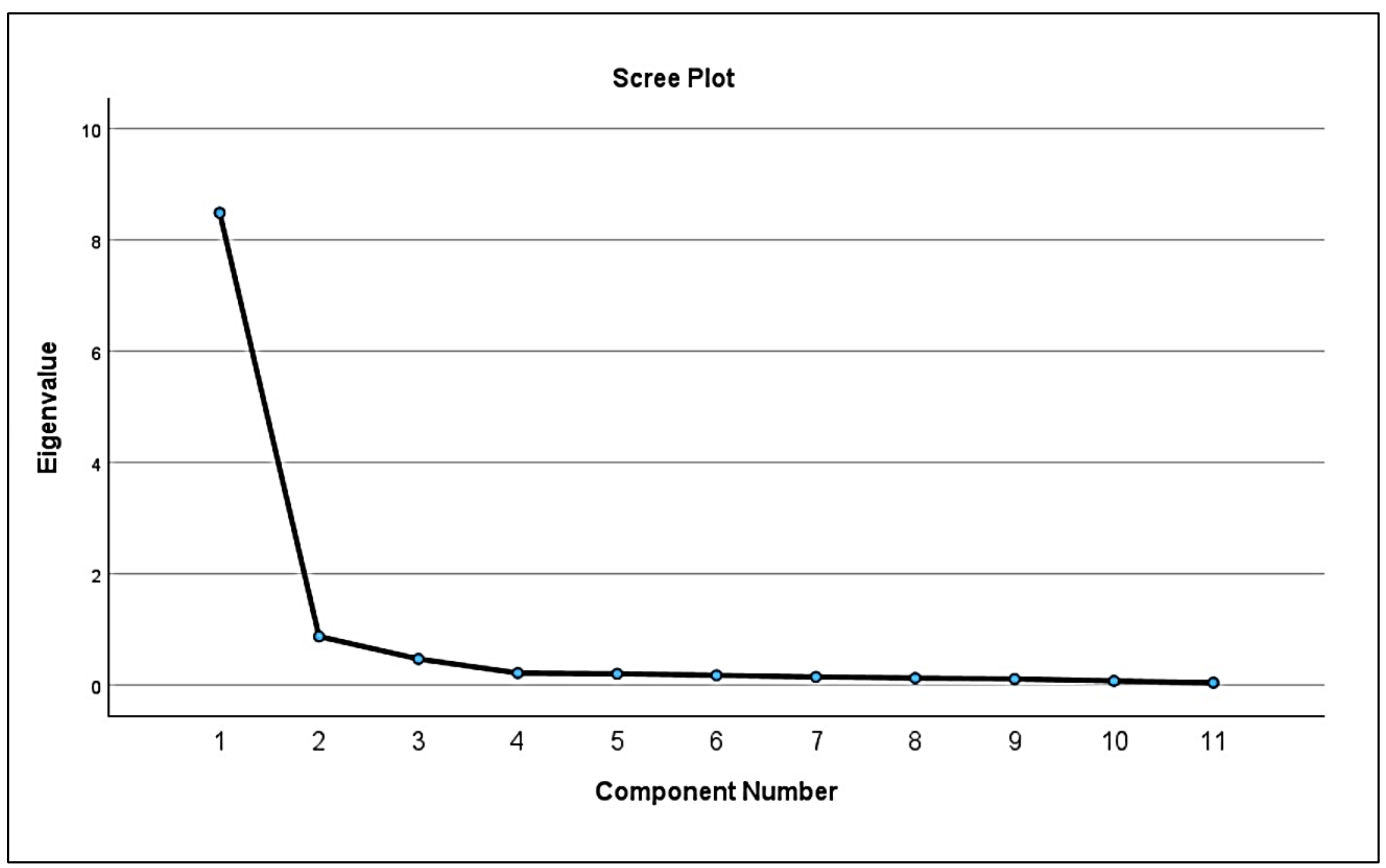

Figure 7. Scree Plot for PCA (Eigenvalues vs. Component Number)

### 4.3.4 Component Loadings

The Component Matrix (Table 10) reveals that all items load strongly on this single extracted factor, with loadings approximately 0.75–0.94. Because only one factor was retained, no rotation was performed.

Table 10. Component Matrix (One Factor Extracted)

| Item | Loading range |
|---|---|
| TI | 0.875–0.903 |
| OR | 0.871–0.937 |
| GE | 0.865–0.874 |
| AIO | 0.811–0.874 |

#### 4.3.5 Interpretation

All 11 items, encompassing technical infrastructure, organizational readiness, governance environment, and AI outcomes, converge onto a single latent dimension of "AI Adoption". This strong unidimensionality indicates that these various aspects of AI capability are highly interrelated for government in the GCC. Given the factor explains over 77% of the variance, it suggests a robust, cohesive measure of overall AI adoption.

### 4.4 Partial Least Squares Structural Equation Modeling (PLS-SEM)

This section presents the Partial Least Squares Structural Equation Modeling (PLS-SEM) results following a two-step analytical approach. First, the Measurement Model is assessed in terms of reflective and formative constructs, examining reliability, convergent validity, and discriminant validity. Second, the Structural Model is evaluated by testing hypothesized path relationships, effect sizes, and overall explanatory power. All computations were performed in SmartPLS 4 using a path weighting scheme with 5,000 bootstrap resamples to ascertain statistical significance.

#### 4.4.1 Measurement Model

##### A. Reflective Constructs

Three constructs, Technical Infrastructure (TI), Organizational Readiness (OR), and AI Outcomes (AIO), were specified as reflective. Table 11 summarizes the indicators' outer loadings, Cronbach's alpha, Composite Reliability (CR), and Average Variance Extracted (AVE).

- **Indicator Loadings.** Every indicator exhibits factor loadings above the recommended threshold of 0.70 (ranging from 0.925 to 0.978), implying strong item reliability.
- **Internal Consistency.** Cronbach's alpha values (0.924–0.960) and Composite Reliability (all > 0.95) indicate excellent internal consistency across the constructs.
- **Convergent Validity.** AVE values (0.868–0.926) surpass the 0.50 benchmark, affirming robust convergent validity.

**Note.** Although collinearity checks signaled a higher VIF for one organizational readiness item (Q7 = 10.813), the item was retained due to its strong theoretical relevance and high loading (0.978). No indication of severe multicollinearity was observed for other indicators.

Table 11. Reflective Constructs: Outer Loadings, Reliability, and Validity (Include columns for each item's loading, Cronbach's alpha, CR, and AVE)

| Construct | **Cronbach's alpha** | **Composite (rho_c)** | **Average variance extracted (AVE)** |
|---|---|---|---|
| TI | 0.949 | 0.967 | 0.908 |
| OR | 0.960 | 0.974 | 0.926 |
| AIO | 0.924 | 0.952 | 0.868 |

##### B. Formative Construct

Table 12 displays outer weights, where Governance Environment (GE) was modeled formatively with two indicators:

- **Q8:** Clear guidelines and regulations

- **Q9:** Ethical and legal standards

Table 12. Outer weights

| | Outer weights |
|---|---|
| Q2 <- TI | 0.366 |
| Q3 <- TI | 0.321 |
| Q4 <- TI | 0.361 |
| Q5 <- OR | 0.362 |
| Q6 <- OR | 0.312 |
| Q7 <- OR | 0.364 |
| Q8 -> GE | 0.655 |
| Q9 -> GE | 0.373 |
| Q10 <- AIO | 0.375 |
| Q11 <- AIO | 0.343 |
| Q12 <- AIO | 0.355 |

- **Outer Weights and Significance.** Q8 (weight = 0.655) and Q9 (weight = 0.373) are both significant ($p < 0.05$), indicating each item contributes uniquely to the Governance Environment (GE) construct.

Collinearity. VIF values for Q8 and Q9 (~4.495) remain below 5, indicating no severe collinearity concerns for a formative specification (see Table 13).

Given these findings, the Governance Environment (GE) is appropriately modelled as a formative construct, where each indicator captures a distinct aspect of the governance environment.

Table 13. Formative Construct (GE): Outer Weights, Loadings, and VIF

| Indicator | Outer Weight (O) | Std. Dev. | t-value | p-value | Outer Loading | VIF |
|---|---|---|---|---|---|---|
| Q8 | 0.655 | 0.144 | 4.564 | 0.000 | 0.984 | 4.495 |
| Q9 | 0.373 | 0.148 | 2.516 | 0.012 | 0.951 | 4.495 |

### C. Discriminant Validity

Discriminant validity was assessed using both the Fornell–Larcker criterion and the HTMT ratio.

- **Fornell–Larcker Criterion.** Each construct's square root of AVE (on the diagonal in Table 14) exceeds its correlations with other constructs, satisfying the criterion.
- **HTMT Ratios.** All HTMT values (Table 15) are below 0.90, further supporting discriminant validity.

Although Technical Infrastructure and Organizational Readiness are highly correlated, they remain sufficiently distinct constructs based on both tests.

Table 14. Fornell–Larcker Criterion (Diagonal entries [√AVE] vs. inter-construct correlations)

| | AIO | TI | OR |
|---|---|---|---|
| AIO | 0.932 | | |
| TI | 0.825 | 0.953 | |
| OR | 0.754 | 0.844 | 0.963 |

Table 15. HTMT Ratios (All ratios < 0.90 cutoff)

| | AIO | TI | OR |
|---|---|---|---|
| AIO | | | |
| TI | 0.877 | | |
| OR | 0.797 | 0.884 | |

### 4.5 Structural Model

After confirming the robustness of the measurement model, the next step entails examining the structural paths. Figure 8 visually depicts the final model, while Table 16 summarizes the path coefficients (β), corresponding t-values, p-values, and 95% confidence intervals (CI). Bootstrapping (5,000 resamples) provides the basis for inferring statistical significance.

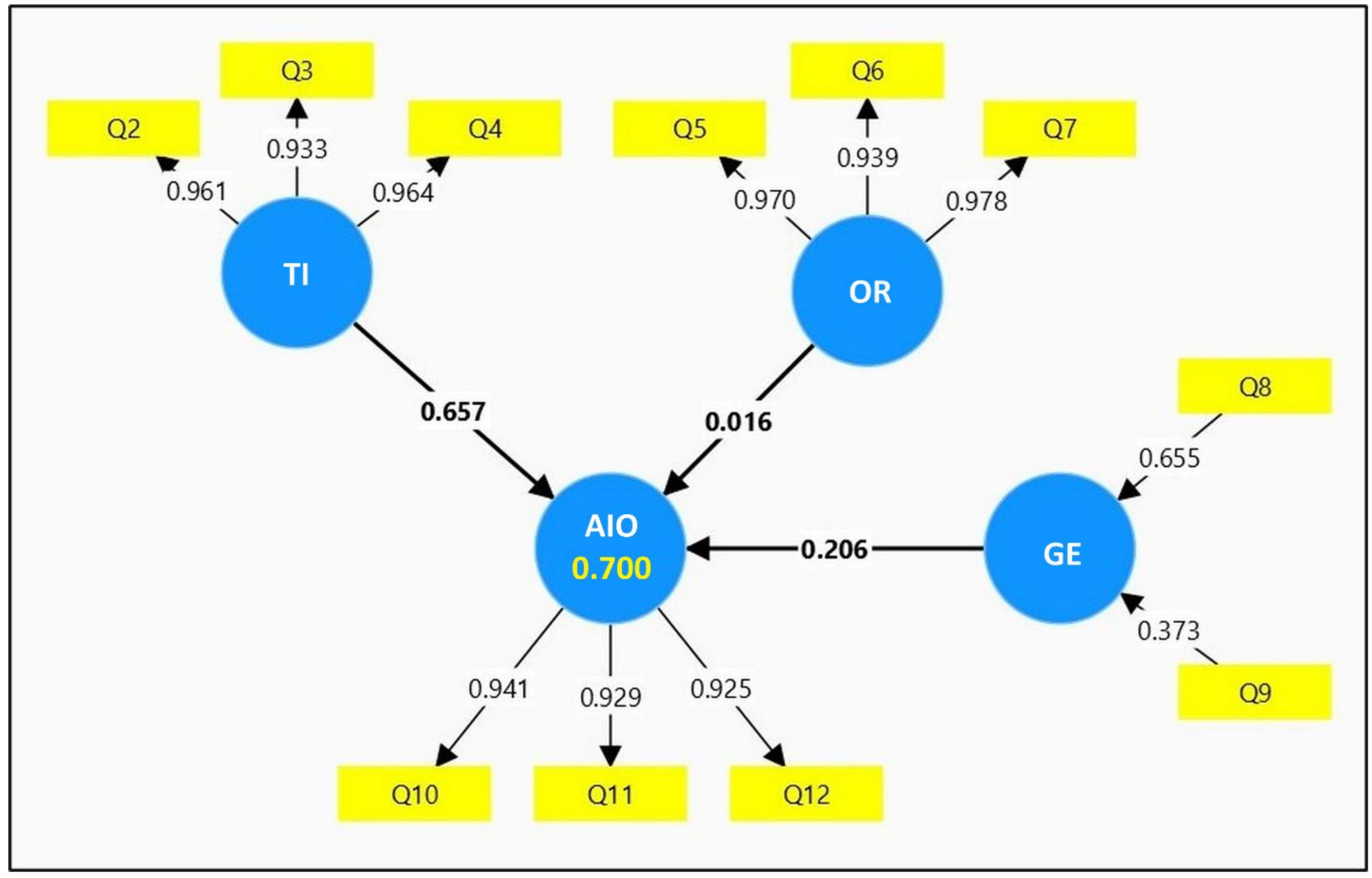

Figure 8. The Final Model

Table 16. Structural Model Path Coefficients and Significance

| Path | β (Original) | t-stat | p-value | 95% CI | Results |
|---|---|---|---|---|---|
| TI → AIO | 0.657 | 8.107 | 0.000 | [0.496, 0.814] | Supported |
| OR → AIO | 0.016 | 0.117 | 0.907 | [−0.252, 0.280] | Not Supported |
| GE → AIO | 0.206 | 2.068 | 0.039 | [0.008, 0.398] | Supported |

#### 4.5.1 Path Analysis and Hypothesis Testing

1. **Technical Infrastructure (TI) → AI Outcomes (AIO)** The results demonstrate a strong, positive, and highly significant influence ($\beta = 0.657$, $t = 8.107$, $p < 0.001$), confirming Hypothesis 1. This underscores the pivotal role of tangible resources, technical capabilities, and support structures in driving successful AI outcomes.
2. **Organizational Readiness (OR) → AI Outcomes (AIO)** Contrary to expectations, organizational readiness shows a small and statistically insignificant path coefficient ($\beta = 0.016$, $p = 0.907$). Hypothesis 2 is thus not supported. A potential explanation may be that once robust technical infrastructure and clear policies are in place, the incremental contribution of broader readiness factors is overshadowed or fully mediated. This non-significance warrants further investigation into how organizational culture, leadership, or skills might interact with technical infrastructure when AI initiatives scale.
3. **Governance Environment (GE) → AI Outcomes (AIO)** Policy displays a moderate but significant effect ($\beta = 0.206$, $t = 2.068$, $p = 0.039$), supporting Hypothesis 3. The finding highlights the importance of clear regulations and ethical guidelines in shaping effective AI adoption.

#### 4.5.2 Coefficient of Determination ($R^2$) and Effect Sizes ($f^2$)

As shown in table 17, Coefficient of Determination ($R^2$), the structural model explains 70% of the variance in AI Outcomes ($R^2 = 0.70$), indicating strong predictive ability. The adjusted $R^2$ value (0.696) remains similar, suggesting model stability. While Effect Sizes ($f^2$), technical infrastructure exhibits a large unique effect ($f^2 = 0.414$), whereas Policy indicates a small yet meaningful effect ($f^2 = 0.029$). Organizational Readiness presents a negligible effect ($f^2 = 0.000$), consistent with its non-significant path.

Table 17. Model Quality

| Dependent Variable | $R^2$ | $R^2$ (Adjusted) |
|---|---|---|
| AI Outcomes | 0.70 | 0.696 |

### 4.6 Proposed AI Adoption Index

Given that a preliminary PCA suggested Technical Infrastructure (TI), Organizational Readiness (OR), and Governance Environment (GE) share a high degree of common variance, and our PLS-SEM indicates that Technical Infrastructure and Governance Environment are the most influential predictors of AI Outcomes, we propose a combined AI Adoption Index. This index incorporates each construct's normalized path coefficient as a weight. Specifically, we begin with the final path coefficients reported in the structural model:

$$\Sigma|\beta| = 0.657 + 0.016 + 0.206 = 0.879$$

To derive weights that sum to 1, we first calculate the total:

$$w_{TI} = \frac{0.657}{0.879} \approx 0.747, \qquad w_{OR} = \frac{0.016}{0.879} \approx 0.018, \qquad w_{GE} = \frac{0.206}{0.879} \approx 0.23$$

Rounded slightly for simplicity, these yield:

$$w_{TI} \approx 0.75, \qquad w_{OR} \approx 0.02, \qquad w_{GE} \approx 0.23$$

Accordingly, we define the Equation for AI Adoption Index. Since the underlying survey items were measured on a 1–5 scale, a multiplier of 20 is applied to normalize the final index to a 0–100 range, as shown in this equation:

$$Index = (w_{TI} \times TI + w_{OR} \times OR + w_{GE} \times GE) \times 20$$

When using the rounded weights, the equation becomes:

$$Index \approx (0.75 \times TI + 0.02 \times OR + 0.23 \times GE) \times 20$$

A resulting index score near 100 indicates an environment highly conducive to AI adoption, while lower scores suggest more limited readiness. Consistent with the PLS-SEM findings, Organizational Readiness receives only a small weight ($\approx$ 0.02%), reflecting its marginal (and statistically non-significant) path coefficient in the structural model. Meanwhile, Technical Infrastructure ($\approx$ 0.75%) and Governance Environment ($\approx$ 0.23%) dominate the index, indicating their comparatively strong roles in driving AI Outcomes.

Researchers or practitioners can adopt these path-based weights directly if they aim to align the index with the empirically derived effects from our model. Alternatively, they may adjust weights if a specific theoretical or policy framework dictates a different emphasis, for instance, elevating the relative importance of adoption factors despite the low coefficient.

### 4.7 Summary of Results

The empirical analyses converge on a clear pattern of AI adoption in GCC public-sector institutions. K-Means clustering revealed two distinct adopter profiles, "High AI Adoption" (predominantly UAE and Saudi Arabia) and "Moderate Adoption" (other GCC states), indicating meaningful heterogeneity in how governments translate national AI ambitions into organizational practice. An exploratory PCA further showed that AI adoption can, at a global level, be treated as a highly cohesive construct: a single dominant component accounts for roughly 77% of the variance across all survey items, suggesting that technical, organizational, governance, and outcome perceptions are strongly intertwined in respondents' views.

PLS-SEM, however, provides a more nuanced decomposition of this overarching construct into three conceptually distinct yet interrelated dimensions: Technical Infrastructure, Organizational Readiness, and Governance Environment. The structural model demonstrates that Technical Infrastructure ($\beta = 0.657$) and Governance Environment ($\beta = 0.206$) are the key drivers of AI Outcomes, together explaining 70% of the variance in perceived AI benefits ($R^2 = 0.70$). In contrast, once these factors are controlled for, Organizational Readiness ($\beta = 0.016$) exerts only a negligible and statistically non-significant effect. This pattern underscores a context in which well-funded infrastructure and clear policy mandates strongly shape AI performance, while internal readiness factors appear less decisive at this early stage of adoption.

Building directly on these findings, the proposed AI Adoption Index translates the final PLS-SEM results into an interpretable 0–100 composite score. The index uses normalized path coefficients as weights, approximately 0.75 for Technical Infrastructure, 0.02 for Organizational Readiness, and 0.23 for Governance Environment, thereby aligning the index with the empirically observed impact of each dimension on AI Outcomes. Higher index scores indicate stronger AI adoption capabilities, driven primarily by robust technical foundations and supportive, ethics-aware governance frameworks, while still retaining a (smaller) contribution from organizational readiness. As such, the index offers a concise, evidence-based metric that can be used to benchmark AI maturity across GCC entities and to track progress over time as infrastructural, organizational, and policy conditions evolve.

## 5. Discussion

By integrating insights from the theory-driven document analysis of related literature and six GCC NASs with data-driven evidence from our survey, PCA, K-Means clustering, and PLS-SEM, this study confirms that Technical Infrastructure and Governance Environment overwhelmingly shape AI Outcomes, together accounting for a large share of the variance. These findings suggest that large-scale national initiatives, bolstered by ample technical infrastructure and clear regulatory directives, can accelerate AI adoption in public-sector agencies. By contrast, Organizational Readiness did not emerge as statistically significant, implying that workforce training, leadership engagement, and cultural acceptance may be subsumed under the broader umbrella of top-down governance and infrastructure-driven rollouts. Nonetheless, such adoption dimensions could become more salient as AI deployments mature beyond pilot phases and demand sustained institutional support. The clustering results further show how respondents perceive AI adoption holistically, revealing a “High AI Adoption” cluster (notably in Saudi Arabia and the UAE) and a “Moderate Adoption” cluster (predominantly other GCC states).

By creating an index that integrates dimensions often omitted in traditional digital maturity evaluations, this work adds nuance to discussions of AI-driven governance in resource-intensive contexts. While existing surveys and indices primarily emphasize broad e-government metrics, they seldom isolate AI-centric factors or address unique governance styles that rely on strong policy directives and generous budget allocations. This study’s focus on technical infrastructure availability, policy clarity, and emerging AI-ethics considerations provides a more context-specific framework, highlighting how well-funded technical investments and centralized directives can spur rapid adoption in early stages of public-sector innovation.

For GCC governments aiming to accelerate AI implementation, the findings suggest that establishing robust technical infrastructure and ensuring clear regulatory mandates are crucial first steps. These factors not only enable rapid piloting of AI projects but also create a conducive environment for scaling. Organizational Readiness, although overshadowed in the current model, may become critical as AI programs mature and require deeper cultural acceptance, continuous skill development, and leadership commitment. The identification of two broad adopter segments, High AI Adoption and Moderate Adoption, points to opportunities for cross-country collaboration, shared best practices, and coordinated policy refinements, which could help less advanced states bridge any adoption gaps.

While the study's focus on IT and AI professionals provides valuable expert insights, it may also skew perceptions of AI adoption toward more favorable views. Future research could incorporate a broader range of perspectives, including frontline government employees and citizens, to gain a more comprehensive understanding of infrastructural availability, policy clarity, and organizational culture. Additionally, the cross-sectional design captures a single point in time, making it challenging to track how AI adoption evolves or to establish causality. Longitudinal studies would allow for a dynamic view of how technical infrastructure, organizational readiness, and governance environment factors interact and shift through different implementation phases. Further qualitative research, such as interviews or focus groups, might deepen exploration into local cultural and religious influences, as well as ethical considerations that underlie trust in AI systems.

By synthesizing theory-driven constructs with data-driven techniques, this study offers a replicable and transparent AI Adoption Index tailored to the GCC context. Its findings underscore the pivotal role of technical infrastructure and policy clarity in driving AI implementations, even as organizational readiness remains a latent factor that may gain prominence as projects move from inception to long-term sustainability. With its utility as both a diagnostic and benchmarking tool, this index can guide policymakers in identifying adoption gaps and strategically allocating resources to foster AI-driven transformation. Moving forward, continued research and iterative refinements of the index will be essential for capturing the evolving interplay among technical infrastructure, regulatory frameworks, organizational dynamics, and cultural norms in shaping AI adoption across the Gulf region.

## 6. Conclusion

This study advances the discourse on AI governance and adoption by introducing a GCC-specific AI Adoption Index built through a theory-driven document analysis, particularly the literature review and NASs analysis, and data-driven statistical validation (survey, clustering, PCA, and PLS-SEM). In contrast to conventional global e-government or digital readiness measures, the resulting index directly reflects the unique structural, policy, and cultural realities of the GCC region. Our findings show that robust technical infrastructure and clear regulatory frameworks are decisive for achieving successful AI outcomes, whereas organizational readiness, often deemed crucial, plays a secondary role once technical infrastructure and policy mandates are in place. These insights underscore how top-down governance and substantial resource investment can shape early-stage AI implementations in resource-intensive settings, though organizational factors may gain prominence in later phases of adoption. Ultimately, the index offers policymakers a nuanced, context-sensitive tool for guiding AI strategies, resource allocation, and collaborative initiatives across the GCC.

The proposed index holds practical value for policymakers and researchers alike, serving as both a diagnostic and benchmarking tool. By mapping specific adoption gaps, it can guide more strategic allocation of resources and capacity-building efforts, especially where AI deployments transition from pilot projects to mature, citizen-centric services. Future work may incorporate longitudinal designs and qualitative methods to capture the evolving interplay of technical infrastructure, governance environment, workforce dynamics, and sociocultural norms, thereby refining this index as AI programs scale and diversify across the GCC. Ultimately, our research underscores the necessity of context-specific frameworks for assessing AI maturity, providing a replicable model that can be adapted for broader regional or global application.

## Acknowledgments

The authors gratefully acknowledge the invaluable support and insights provided by the government officials, policymakers, and academic experts in the GCC who participated in this study. Finally, we appreciate the constructive feedback from the anonymous reviewers, which has significantly contributed to improving the clarity and rigor of this manuscript.